\documentstyle[12pt]{article} \voffset=0mm \headheight=0mm
\date{}
\author{Valerii Dryuma\thanks{Work supported in part by Grant RFFI, Russia-Moldova}\\[5mm]
{\it Institute of Mathematics and Informatics, AS RM,}\\[3mm] {\it
5 Academiei Street, 2028 Kishinev, Moldova},\\[3mm]{\it e-mail:
valery@dryuma.com; valdryum@gmail.com} }
\title{On equations defining the Ricci flows of manifolds }
\newtheorem{theorem}{Theorem}
\newtheorem{pr}{Proposition}

\begin{document}
\maketitle
\date{}
\maketitle
\begin{abstract}
\ \ \ \ The examples of the Ricci flows of the four-dimensional manifolds which are
determined by help of nonlinear differential equations of the type of Monge-Ampere are constructed.Their particular   solutions are derived and their properties are discussed.
\end{abstract}


\medskip
\section{$4D$ model of the Ricci-flow }

  We study the system of equations 
  \begin{equation}\label{ricci}
\frac{\partial}{\partial t}g_{ij}(\vec x,t)=-2 R_{ij}(g)
 \end{equation}
describing the Ricci flows of the four dimensional manifolds which are  endowed  by the  metrics 
of the form
  \begin{equation}\label{dr:eq2} {{\it
ds}}^{2}=A(x,y,t){{\it du}}^{2}+2\,B(x,y,t){\it du}\,{\it dv}+{
\it du}\,{\it dx}+C(x,y,t){{\it dv}}^{2}+{\it dv}\,{\it dy},
\end{equation}
where the components of metrics are dependent from two coordinates $(x,y)$ and from the parameter $t$.

    In this case the Ricci-tensor of the metric
     $(\ref{ricci})$
     has a five components
$$ R_{uu},~ R_{uv},~ R_{vv},~
 R_{ux}, ~R_{vx}.
 $$

From the conditions of compatibility of the equations
(\ref{ricci}) we find that
    they are reduced to the equation
\begin{equation}\label{dr:eq3}
4\,\left ({\frac {\partial ^{2}}{\partial {x}^{2}}}h(x,y,t)\right
){ \frac {\partial ^{2}}{\partial {y}^{2}}}h(x,y,t)-4\,\left
({\frac {
\partial ^{2}}{\partial x\partial y}}h(x,y,t)\right )^{2}-{\frac {
\partial }{\partial t}}h(x,y,t)=0
\end{equation}
and components of the metric (\ref{dr:eq2}) take the form
\begin{equation}\label{dr:eq4}
  B(x,y,t)={\frac
{\partial ^{2}}{\partial x\partial y}}h(x,y,t),~ C(x,y,t)=-{\frac
{\partial ^{2}}{\partial {x}^{2}}}h(x,y,t),~ A(x,y,t)=-{\frac
{\partial ^{2}}{\partial {y}^{2}}}h(x,y,t).
\end{equation}

     Thus the study of the properties of the Ricci flow on a 4-dimensional manifold
     with the metric (\ref{dr:eq2}) with the coefficients (\ref{dr:eq4}) lead
     to integration of the equation (\ref{dr:eq3}). Nyis equation is the first order relatively 
the parameter $\tau$ and has form of the Monge-Ampere equation relatively of the variables $x,y$ 

\section{Particular solutions}

    Let us consider some elementary solutions of equation (\ref{dr:eq3})

   After the substitution of the form
$$
h(x,y,t)=H\left({\frac {x}{y}}-t\right){y}^{4}
$$
the equation (\ref{dr:eq3}) takes the form
\begin{equation}\label{dr:eq5}
48\,\left (D^{\left (2\right )}\right )(H)(\eta)H(\eta)-36\,\left
( \mbox {D}(H)(\eta)\right )^{2}+\mbox {D}(H)(\eta)=0,
\end{equation}
where
$$
\eta={\frac {x}{y}}-t.
$$

   Solution of the equation (\ref{dr:eq4}) may be present in form
$$
   \eta(H)={\it \_C2}-144\,{\frac {\sqrt [4]{H}}{{K}^{3}}}+$$$$\left(-48\,
\ln (1-\sqrt [4]{H}K)+24\,\ln (1+\sqrt [4]{H}K+\sqrt
{H}{K}^{2})+48\, \sqrt {3}\arctan({\frac {\sqrt {3}\sqrt
[4]{H}K}{2+\sqrt [4]{H}K}}) \right){K}^{-4},
 $$
where $K$ and ${\it \_C2}$ are constants.

    Remark that if the value $K=-1$ the function $ \eta(H)$ has a
    break.

    The simplest singular solution of (\ref{dr:eq4}) is
$$ h(x,y,t)= $$ $$=\frac{\left(6\,{\frac {{B}^{2}{x}^{4}}{k}}+{\it
\_C1}\,\sqrt {x} \cos(1/2\,\sqrt {23}\ln (x))+{\it \_C2}\,\sqrt
{x}\sin(1/2\,\sqrt {23} \ln
(x))+yB{x}^{3}+1/24\,{y}^{2}k{x}^{2}\right)}{\left (2\,kt+{\it
\_C4} \right )^{-1}}.
$$

    More complicated solution has the form
    $$
h(x,y,t)=-1/2\,{\frac {{{\it \_C1}}^{2}\mu\,\sqrt {{\it
\_C1}\,{\it \_c}_{{1}}}\sqrt {2}\tan(1/2\,\sqrt {{\it \_C1}\,{\it
\_c}_{{1}}} \left (x-t\right )\sqrt {2})}{{{\it
\_c}_{{1}}}^{2}}}+{\frac {\mu\,{ \it \_C1}\,y}{{\it
\_c}_{{1}}}}-$$$$-1/4\,{\frac {\mu\,\sqrt {{\it \_C1}\,{ \it
\_c}_{{1}}}\sqrt {2}{y}^{2}}{{\it \_c}_{{1}}\tan(1/2\,\sqrt {{\it
\_C1}\,{\it \_c}_{{1}}}\left (x-t\right )\sqrt {2}){\it \_C1}}}.
 $$

    A more general solutions of the equation (\ref{dr:eq4}) are derived by
    the method of parametric representations of functions and their
    derivatives [1-2].

    Let us apply its to the equation (\ref{dr:eq4}).

      After the change of the variables, the function and its derivative
\begin{equation} \label{Dr11}
h(x,y,t) \rightarrow u(x,\tau,t),~ y \rightarrow v(x,\tau,t),~ h_x
\rightarrow u_x-\frac{u_{\tau}}{v_{\tau}}v_x=p,~ h_t \rightarrow
u_t-\frac{u_{\tau}}{v_{\tau}}v_t=q,$$$$ h_y \rightarrow
\frac{u_{\tau}}{v_{\tau}}=r, ~ h_{xx} \rightarrow
p_x-\frac{p_{\tau}}{v_{\tau}}v_x,~h_{xy}\rightarrow
\frac{p_{\tau}}{v_{\tau}}=q_x-\frac{q_{\tau}}{v_{\tau}}v_x,~h_{yy}\rightarrow
\frac{r_{\tau}}{v_{\tau}}
\end{equation}
 we derive from the equation
 \begin{equation} \label{Dr12}
\left ({\frac {\partial ^{2}}{\partial {x}^{2}}}h(x,y,t)\right ){
\frac {\partial ^{2}}{\partial {y}^{2}}}h(x,y,t)-\left ({\frac {
\partial ^{2}}{\partial x\partial y}}h(x,y,t)\right )^{2}-{\frac {
\partial }{\partial t}}h(x,y,t)=0
 \end{equation}
the relation between the functions $u(x,\tau,t)$ and
$v(x,\tau,t)$, where variable $\tau$ is considered as parameter
 \begin{equation} \label{Dr13}
-\left ({\frac {
\partial ^{2}}{\partial {x}^{2}}}u(x,\tau,t)\right )\left ({\frac {
\partial }{\partial \tau}}v(x,\tau,t)\right )\left ({\frac {\partial }
{\partial \tau}}u(x,\tau,t)\right ){\frac {\partial ^{2}}{\partial
{ \tau}^{2}}}v(x,\tau,t)-$$$$-\left ({\frac {\partial }{\partial
\tau}}u(x, \tau,t)\right )\left ({\frac {\partial ^{2}}{\partial
{x}^{2}}}v(x, \tau,t)\right )\left ({\frac {\partial }{\partial
\tau}}v(x,\tau,t) \right ){\frac {\partial ^{2}}{\partial
{\tau}^{2}}}u(x,\tau,t)+$$$$+\left ({\frac {\partial }{\partial
\tau}}u(x,\tau,t)\right )^{2}\left ({ \frac {\partial
^{2}}{\partial {x}^{2}}}v(x,\tau,t)\right ){\frac {
\partial ^{2}}{\partial {\tau}^{2}}}v(x,\tau,t)-$$$$-\left ({\frac {
\partial ^{2}}{\partial \tau\partial x}}u(x,\tau,t)\right )^{2}\left (
{\frac {\partial }{\partial \tau}}v(x,\tau,t)\right )^{2}+\left
({\frac {\partial ^{2}}{\partial {x}^{2}}}u(x,\tau,t)\right )
\left ({\frac {\partial }{\partial \tau}}v(x,\tau,t)\right )^{2}{
\frac {\partial ^{2}}{\partial
{\tau}^{2}}}u(x,\tau,t)+$$$$+2\,\left ({ \frac {\partial
^{2}}{\partial \tau\partial x}}u(x,\tau,t)\right ) \left ({\frac
{\partial }{\partial \tau}}v(x,\tau,t)\right )\left ({ \frac
{\partial }{\partial \tau}}u(x,\tau,t)\right ){\frac {\partial ^
{2}}{\partial \tau\partial x}}v(x,\tau,t)-$$$$-\left ({\frac
{\partial }{
\partial \tau}}u(x,\tau,t)\right )^{2}\left ({\frac {\partial ^{2}}{
\partial \tau\partial x}}v(x,\tau,t)\right )^{2}-\left ({\frac {
\partial }{\partial t}}u(x,\tau,t)\right )\left ({\frac {\partial }{
\partial \tau}}v(x,\tau,t)\right )^{4}+$$$$+\left ({\frac {\partial }{
\partial \tau}}u(x,\tau,t)\right )\left ({\frac {\partial }{\partial t
}}v(x,\tau,t)\right )\left ({\frac {\partial }{\partial
\tau}}v(x,\tau ,t)\right )^{3} =0.
 \end{equation}

    Note  that the relation (\ref{Dr13}) under the condition $v(x,\tau,t)=\tau$
    is equivalent to the equation
 $$
{\frac {\partial ^{2}}{\partial {x}^{2}}}u(x,t,\tau){ \frac
{\partial ^{2}}{\partial {t}^{2}}}u(x,t,\tau)-\left ({\frac {
\partial ^{2}}{\partial t\partial x}}u(x,t,\tau)\right )^{2}-{\frac {
\partial }{\partial \tau}}u(x,t,\tau)=0
 $$
similar to equation (\ref{Dr12}) and transformed into an equation
of more general form if the functions $u(x,\tau,t)$,
$(v(x,\tau,t)$ are dependent.

    For example under the condition
$$ u(x,\tau,t)=\tau\,{\frac {\partial }{\partial
\tau}}\omega(x,\tau,t)- \omega(x,\tau,t),~ v(x,\tau,t)={\frac
{\partial }{\partial \tau}}\omega(x,\tau,t)
 $$
from the (\ref{Dr13}) we obtain the p.d.e. for the function
$\omega(x,\tau,t)$
 \begin{equation} \label{Dr14} -{\frac {\partial ^{2}}{\partial
{x}^{2}}}\omega(x,\tau,t)+\left ({ \frac {\partial ^{2}}{\partial
{\tau}^{2}}}\omega(x,\tau,t)\right ){ \frac {\partial }{\partial
t}}\omega(x,\tau,t)=0.
 \end{equation}

   The other condition
 $$ v(x,\tau,t)=\tau\,{\frac {\partial
}{\partial \tau}}\omega(x,\tau,t)- \omega(x,\tau,t),~
u(x,\tau,t)={\frac {\partial }{\partial \tau}}\omega(x,\tau,t) $$
lead to the equation
 \begin{equation} \label{Dr15}
{\frac {\partial ^{2}}{\partial {x}^{2}}}\omega(x,\tau,t)+\left ({
\frac {\partial ^{2}}{\partial {\tau}^{2}}}\omega(x,\tau,t)\right
){ \tau}^{3}{\frac {\partial }{\partial t}}\omega(x,\tau,t)=0.
\end{equation}

    Let us consider some solutions of the equation (\ref{Dr14}).

Under the substitution
 $$
\omega(x,\tau,t)=A(x-kt,\tau)
 $$
it is reduced at the equation
\begin{equation} \label{Dr16}
{\frac {\partial ^{2}}{\partial {\eta}^{2}}}A(\eta,\tau)+k\left ({
\frac {\partial ^{2}}{\partial {\tau}^{2}}}A(\eta,\tau)\right
){\frac {\partial }{\partial \eta}}A(\eta,\tau) =0
\end{equation}
where $\eta=x-k t$.

   This equation is reduced after the  Legendre transformation to the Euler-Trikomi equation
   and therefore its solutions depend radically on the sign of the parameter $k$.

    Each solution (\ref{Dr16}) corresponds to the solution of the
    original equation (\ref{Dr13}) which is obtained by eliminating the parameter $\tau$
    from the correlations
\begin{equation} \label{Dr17} y-\tau\,{\frac {\partial
}{\partial \tau}}\omega(x,\tau,t)+ \omega(x,\tau,t)=0,~
h(x,y,t)-{\frac {\partial }{\partial \tau}}\omega(x,\tau,t)=0.
\end{equation}

   Here is an example of.

\section{Finite-dimensional reduction}

   After the substitutions
$$ A(x,y,t)={\it A_0}(t)+{\it A_1}(t)x+{\it A_2}(t)y+{\it
A_3}(t){x}^{2}+{ \it A_4}(t)yx+{\it A_5}(t){y}^{2},$$$$
B(x,y,t)={\it B_0}(t)+{\it B_1}(t)x+{\it B_2}(t)y+{\it
B_3}(t){x}^{2}+{ \it B_4}(t)xy+{\it B_5}(t){y}^{2},$$$$
C(x,y,t)={\it C_0}(t)+{\it C_1}(t)x+{\it C_2}(t)y+{\it
C_3}(t){x}^{2}+{ \it C_4}(t)xy+{\it C_5}(t){y}^{2},$$$$ {\it
B_4}(t)=-2\,{\it A_3}(t),~ {\it A_4}(t)=-2\,{\it B_5}(t),~ {\it
C_4}(t)=-2\,{\it B_3}(t),~ {\it C_5}(t)={\it A_3}(t)
 $$
the metric (\ref{dr:eq2}) takes the form
\begin{equation}\label{metr2}
  {{\it ds}}^{2}=\left ({\it A_0}(t )+{\it
A_1}(t)x+{\it A_2}(t)y+{\it A_3}(t){x}^{2}-2\,{\it B_5}(t)yx+{\it
A_5}(t){y}^{2}\right ){{\it du}}^{2}+$$$$+2\,\left ({\it
B_0}(t)+{\it B_1}(t)x +{\it B_2}(t)y+{\it B_3}(t){x}^{2}-2\,{\it
A_3}(t)xy+{\it B_5}(t){y}^{2} \right ){\it du}\,{\it
dv}+$$$$+\left ({\it C_0}(t)+{\it C_1}(t)x+{\it C_2}(t) y+{\it
C_3}(t){x}^{2}-2\,{\it B_3}(t)xy+{\it A_3}(t){y}^{2}\right ){{\it
dv}}^{2}+{\it dx}\,{\it du}+{\it dy}\,{\it dv}
\end{equation}
 and from the system of equations (\ref{ricci}) is obtained system of ODE's
\begin{equation}\label{sys}
  {\frac {d}{dt}}{\it B_3}(t)=24\,{\it B_3}(t){\it
A_3}(t)-24\,{\it C_3}(t){ \it B_5}(t),$$$$ {\frac {d}{dt}}{\it
C_3}(t)=-48\,{\it C_3}(t){\it A_3}(t)+48\,\left ({ \it
B_3}(t)\right )^{2},$$$$ {\frac {d}{dt}}{\it A_3}(t)=-8\,{\it
C_3}(t){\it A_5}(t)+24\,\left ({\it A_3}(t)\right )^{2}-16\,{\it
B_3}(t){\it B_5}(t),$$$$ {\frac {d}{dt}}{\it B_5}(t)=24\,{\it
A_3}(t){\it B_5}(t)-24\,{\it B_3}(t){ \it A_5}(t),$$$$ {\frac
{d}{dt}}{\it A_5}(t)=-48\,{\it A_5}(t){\it A_3}(t)+48\,\left ({
\it B_5}(t)\right )^{2}
\end{equation}
on the functions $A_3(t),~B_3(t),~C_3(t),~A_5(t),~B_5(t)$ from
which expressed the rest coefficients of the metric (\ref{metr2}).

   The system of equations (\ref{sys}) has a first integral $M$ of the
   type
   $$
{\it C_3}(t){\it B_5}(t)^{2}+M{\it C_3}(t)^ {2}{\it B_5}(t)-{\it
A_5}(t){\it B_3}(t)^{2}+2\,M{ \it B_3}(t)^{3}-3\,M{\it C_3}(t){\it
A_3}(t){\it B_3}(t) =0
 $$
with the help of which the order can be reduced.

 In result we get the system with parameter
 $$
{\frac {d}{dt}}{\it B_3}(t)=24\,{\it B_3}(t){\it A_3}(t)-24\,{\it
C_3}(t){ \it B_5}(t),\quad {\frac {d}{dt}}{\it C_3}(t)=-48\,{\it
C_3}(t){\it A_3}(t)+48\,\left ({ \it B_3}(t)\right )^{2},$$$${\it
B_3}(t)^{2}{\frac {d}{dt}}{\it A_3}(t)=-8\,{ \it C_3}(t)^{2}{\it
B_5}(t)^{2}-8\,M {\it C_3}(t)^{3}{\it B_5}(t)-16\,{\it C_3}(t)M
{\it B_3}(t)^{3}+$$$$+24\,M{\it C_3}(t)^{2}{\it A_3}(t){\it
B_3}(t) +24{\it A_3}(t)^{2}{\it B_3}(t)^{2}-16\, {\it B_5}(t){\it
B_3}(t)^{3} , $$~ $${\it B_3}(t){\frac {d}{dt}}{\it
B_5}(t)=24\,{\it A_3}(t){\it B_5}(t){\it B_3}(t)-24\,{\it
C_3}(t){\it B_5}(t)^{2}-24\,M{\it C_3}(t)^{2}{\it
B_5}(t)-$$$$-48\,M{\it B_3}(t)^{3}+72\, M{\it C_3}(t){\it
A_3}(t){\it B_3}(t).
 $$
\section{Monge-Ampere flows}

\begin{theorem}    The flows of Ricci of the $4D$- manifolds which are endowed  by the metric
with local coordinates $(x,y,z,t)$ the components of which are dependent from two coordinates $(x,t)$ and from the parameter $\tau$ 
    $$
    {{\it ds}}^{2}=\left ({\frac {\partial ^{2}}{\partial {x}^{2}}}f(x,t,
\tau)\right ){{\it dx}}^{2}+2\,\left ({\frac {\partial
^{2}}{\partial t\partial x}}f(x,t,\tau)\right ){\it dx}\,{\it
dt}+\left ({\frac {
\partial ^{2}}{\partial {t}^{2}}}f(x,t,\tau)\right ){{\it dy}}^{2}+$$$$+2\,
\left ({\frac {\partial ^{2}}{\partial t\partial
x}}f(x,t,\tau)\right ){\it dy}\,{\it dz}+\left ({\frac {\partial
^{2}}{\partial {x}^{2}}}f( x,t,\tau)\right ){{\it dz}}^{2}+\left
({\frac {\partial ^{2}}{
\partial {t}^{2}}}f(x,t,\tau)\right ){{\it dt}}^{2}
$$
 is defined by the equation
\begin{equation} \label{Dr18}{\frac {\partial }{\partial \tau}}f(x,t,\tau)=\ln \left({\frac {
\partial ^{2}}{\partial {t}^{2}}}f(x,t,\tau){\frac {\partial ^
{2}}{\partial {x}^{2}}}f(x,t,\tau)-\left ({\frac {\partial ^{2}}{
\partial t\partial x}}f(x,t,\tau)\right )^{2}\right)
\end{equation}
 \end{theorem}

    Let us consider some solutions of the equation (\ref{Dr18}).

     With aim of convenience we rewrite if in the form
\begin{equation} \label{Dr19}
     \left ({\frac {\partial ^{2}}{\partial {x}^{2}}}f(x,y,\tau)\right ){
\frac {\partial ^{2}}{\partial {y}^{2}}}f(x,y,\tau)-\left ({\frac
{
\partial ^{2}}{\partial x\partial y}}f(x,y,\tau)\right )^{2}-{e^{{
\frac {\partial }{\partial \tau}}f(x,y,\tau)}}=0,
 \end{equation}
where the variable $t$ is changed on $y$.

    After the $(u,v)$-transformation with condition
    $u(x,t,\tau)=t$ the equation (\ref{Dr19}) takes the form
    $$
    \left ({\frac {\partial ^{2}}{\partial {x}^{2}}}v(x,t,\tau)\right ){
\frac {\partial ^{2}}{\partial {t}^{2}}}v(x,t,\tau)-\left ({\frac
{
\partial ^{2}}{\partial t\partial x}}v(x,t,\tau)\right )^{2}-{e^{-{
\frac {{\frac {\partial }{\partial \tau}}v(x,t,\tau)}{{\frac {
\partial }{\partial t}}v(x,t,\tau)}}}}\left ({\frac {\partial }{
\partial t}}v(x,t,\tau)\right )^{4}=0.
$$

    Particular solutions of this equation are obtained with the help of
    additional conditions.
    For example in the case
    $$
    {\frac {\partial }{\partial \tau}}v(x,t,\tau)={\frac {\partial }{
\partial t}}v(x,t,\tau)
$$ we find that the function $v(x,t,\tau)$ has the form $$
v(x,t,\tau)={\it \_F1}(x,\tau+t)=h(x,\eta),
 $$
where the function $h(x,\eta)$
satisfies the equation
\begin{equation} \label{Dr20}\left ({\frac {\partial ^{2}}{\partial {\eta}^{2}}}h(x,\eta)\right
){ \frac {\partial ^{2}}{\partial {x}^{2}}}h(x,\eta)-\left ({\frac
{
\partial ^{2}}{\partial \eta\partial x}}h(x,\eta)\right )^{2}-{e^{-1}}
\left ({\frac {\partial }{\partial \eta}}h(x,\eta)\right )^{4} =0.
\end{equation}

After the $(u,v)$-transformation with the conditions
 $$
 v(x,t)=t{\frac {\partial }{\partial
t}}\omega(x,t)-\omega(x,t),~ u(x,t)={\frac {\partial }{\partial
t}}\omega(x,t)
$$
   the equation (\ref{Dr20}) is reduced to the linear equation
   $$
   {\frac {\partial ^{2}}{\partial {x}^{2}}}\omega(x,t)+\left ({\frac {
\partial ^{2}}{\partial {t}^{2}}}\omega(x,t)\right ){e^{-1}}=0$$
with general solution
 $$
 \omega(x,t)={\it \_F1}(t-i\sqrt
{{e^{-1}}}x)+{\it \_F2}(t+i\sqrt {{e^ {-1}}}x)
 $$
containing two arbitrary functions.

   With the help of the function $\omega(x,t)$ we can obtain
  large  class of solutions of the equation (\ref{Dr19})
   in parametric form.

    For example, in the case
    $$
    {\it \_F1}(t-i\sqrt {{e^{-1}}}x)=\cosh(+-i\sqrt {{e^{-1}}}x)
{\it \_F2}(t+i\sqrt {{e^{-1}}}x)=\sinh(t+i\sqrt {{e^{-1}}}x) $$ we
get
 $$
  \omega(x,t)=\cosh(t)\cos(\sqrt
{{e^{-1}}}x)+\sinh(t)\cos(\sqrt {{e^{-1 }}}x) $$ and elimination
of the parameter $t$ from the relations $$
\eta-t\cos({e^{-1/2}}x)\cosh(t)-t\sinh(t)\cos({e^{-1/2}}x)+\cosh(t)
\cos({e^{-1/2}}x)+\sinh(t)\cos({e^{-1/2}}x)=0,$$$$
h(x,\eta)-\cosh(t)\cos({e^{-1/2}}x)-\sinh(t)\cos({e^{-1/2}}x)
 $$
lead to the function
$$
 h(x,\eta)={e^{{\it LambertW}({\frac
{\eta\,{e^{-1}}}{\cos({e^{-1/2}}x) }})+1}}\cos({e^{-1/2}}x)
 $$
which is solution of the equation (\ref{Dr20}).

With the help of solutions of the equation (\ref{Dr20}) can be
constructed solutions of the equation (\ref{Dr19}).

\section{More general solution}

     The equation (\ref{Dr19}) after the $(u,v)$-transformation
     with conditions
     $$
     -{\frac {-\left ({\frac {\partial }{\partial \tau}}u(x,t,\tau)\right )
{\frac {\partial }{\partial t}}v(x,t,\tau)+\left ({\frac {\partial
}{
\partial t}}u(x,t,\tau)\right ){\frac {\partial }{\partial \tau}}v(x,t
,\tau)}{{\frac {\partial }{\partial t}}v(x,t,\tau)}}=Ax,$$$$
v(x,t,\tau)={\it \_F1}(x,u(x,t,\tau)-A x \tau $$
 takes the form
\begin{equation} \label{Dr21} \left ({\frac {\partial ^{2}}{\partial
{\eta}^{2}}}h(x,\eta)\right ){ \frac {\partial ^{2}}{\partial
{x}^{2}}}h(x,\eta)-\left ({\frac {
\partial ^{2}}{\partial \eta\partial x}}h(x,\eta)\right )^{2}-{e^{Ax}}
\left ({\frac {\partial }{\partial \eta}}h(x,\eta)\right )^{4} =0,
\end{equation}
where
 $$ \eta=u(x,t,\tau)-Ax\tau,
  $$
   $$ h(x,\eta)={\it
\_F1}(x,u(x,t,\tau)-A x \tau).
 $$

   It is reduced to the equation
   \begin{equation} \label{Dr22}
   {e^{Ax}}{\frac {\partial ^{2}}{\partial {t}^{2}}}\omega(x,t)+{\frac {
\partial ^{2}}{\partial {x}^{2}}}\omega(x,t)=0
\end{equation}
 after the $(u,v)$-transformation with conditions $$
v(x,t)=t{\frac {\partial }{\partial t}}\omega(x,t)-\omega(x,t),~
 u(x,t)={\frac {\partial }{\partial
t}}\omega(x,t).
 $$

    Simplest solution of the equation (\ref{Dr22}) has the form
    $$
    \omega(x,t)=\left ({\it \_C1}\,\cos(\sqrt {{\it \_c}_{{1}}}t)+{\it
\_C2}\,\sin(\sqrt {{\it \_c}_{{1}}}t)\right )\cdot$$$$\cdot\left
({\it \_C3}\,{\it BesselJ}(0,2\,\sqrt {-{\frac {{\it
\_c}_{{1}}}{{A}^{2}}}}\sqrt {{e^{Ax }}})+{\it \_C4}\,{\it
BesselY}(0,2\,\sqrt {-{\frac {{\it \_c}_{{1}}}{{ A}^{2}}}}\sqrt
{{e^{Ax}}})\right )
 $$
and elimination of the parameter $t$ from the relations
 $$
\eta+t\sin(\sqrt {{\it \_c}_{{1}}}t)\sqrt {{\it \_c}_{{1}}}{\it
BesselJ}(0,2\,\sqrt {-{\it \_c}_{{1}}}\sqrt {{e^{x}}})+\cos(\sqrt
{{ \it \_c}_{{1}}}t){\it BesselJ}(0,2\,\sqrt {-{\it
\_c}_{{1}}}\sqrt {{e^ {x}}})=0,$$$$ h(x,\eta)+\sin(\sqrt {{\it
\_c}_{{1}}}t)\sqrt {{\it \_c}_{{1}}}{\it BesselJ}(0,2\,\sqrt
{-{\it \_c}_{{1}}}\sqrt {{e^{x}}})=0
 $$
 gives solution of the equation (\ref{Dr21}) defined from the
relation $$
 \eta\,\sqrt {{\it \_c}_{{1}}}+\arcsin\left({\frac
{h(x,\eta)}{\sqrt {{\it \_c}_{{1}}}{\it BesselJ}(0,2\,\sqrt {-{\it
\_c}_{{1}}}\sqrt {{e^{x}}}) }}\right)h(x,\eta)+$$$$+\sqrt {{\it
\_c}_{{1}}\left ({\it BesselJ}(0,2\,\sqrt {-{ \it \_c}_{{1}}}\sqrt
{{e^{x}}})\right )^{2}-\left (h(x,\eta)\right )^{ 2}} =0, $$
 (at the conditions $A=1$ and  $\it \_C2=0,\it \_C4=0,\it \_C1=1$).

    The solution of the equation (\ref{Dr19}) which corresponds the function $h(x,\eta)$
    is determined from the relation
$$
 y-h\left(x,f(x,y,\tau)-x \tau\right)=0
$$
and can be very complicated.

{\small
\centerline{\bf References:}

\smallskip
\noindent 1. V. Dryuma, {\it On nonlinear equations associated with developable, ruled and minimal surfaces}.\\ ArXiv:1002.0952 v1 [physycs.gen-ph] 4 Feb.2010 p. 1-13.

\smallskip
\noindent 2. V. Dryuma, {\it The Riemann and Einsten-Weyl
geometries in theory of differential equations, their applications
and all that}. A.B.Shabat et all.(eds.), New Trends in
Integrability and Partial Solvability, Kluwer Academic Publishers,
Printed in the Netherlands , 2004, p.115--156.}
\end{document}